\documentclass{jcgt}

\setciteauthor{Russell A. Brown}
\setcitetitle{A Comparison of Two Dynamic k-d Trees}

\accepted{submitted}{accepted}{published}{Editor Name}{vol}{issue}{1}{1}{year}
\seturl{http://jcgt.org/published/vol/issue/num/}

\usepackage[usenames,dvipsnames]{xcolor}
\usepackage{tcolorbox}
\usepackage{tabularx}
\usepackage{array}
\usepackage{tabto}
\usepackage{colortbl}
\tcbuselibrary{skins}
\usepackage{mdframed}

\newcolumntype{Y}{>{\raggedleft\arraybackslash}X}
\tcbset{tab1/.style={fonttitle=\bfseries\large,fontupper=\normalsize\sffamily,
colback=yellow!10!white,colframe=red!75!black,colbacktitle=Salmon!40!white,
coltitle=black,center title,freelance,frame code={
\foreach \n in {north east,north west,south east,south west}
{\path [fill=red!75!black] (interior.\n) circle (3mm); };},}}

\tcbset{tab2/.style={enhanced,fonttitle=\bfseries,fontupper=\normalsize\sffamily,
colback=white!10!white,colframe=black!50!black,colbacktitle=Salmon!40!white,
coltitle=black,center title}}

\usepackage{listings}

\usepackage{xcolor}

\definecolor{codegreen}{rgb}{0,0.6,0}
\definecolor{codegray}{rgb}{0.5,0.5,0.5}
\definecolor{codepurple}{rgb}{0.58,0,0.82}
\definecolor{backcolour}{rgb}{0.99,0.98,0.98}

\lstdefinestyle{mystyle}{
    backgroundcolor=\color{backcolour},   
    commentstyle=\color{codegreen},
    keywordstyle=\color{magenta},
    numberstyle=\tiny\color{codegray},
    stringstyle=\color{codepurple},
    basicstyle=\ttfamily\footnotesize,
    breakatwhitespace=true,         
    breaklines=true,                 
    captionpos=b,                    
    keepspaces=true,                 
    numbers=left,                    
    numbersep=5pt,                  
    showspaces=false,                
    showstringspaces=false,
    showtabs=false,                  
    tabsize=2
}

\begin{document}

\title{A Comparison of Two Dynamic k-d Trees}

\author
       {Russell A. Brown}

\maketitle

\begin{abstract}
\small

Two methods have been proposed for building and modifying a dynamic \emph{k}-d tree. One method stores the dynamic tree as a single \emph{k}-d tree and rebalances that tree by rebuilding subtrees within the tree when those subtrees become unbalanced due to insertion of a \emph{k}-dimensional tuple into the tree or deletion of a tuple from the tree. A second method composes a dynamic tree as a set of static \emph{k}-d trees whose sizes are increasing integer powers of two; this tree's balance is maintained by rebuilding a static tree within the set upon insertion or deletion of a tuple. This article describes insertion and deletion algorithms for the second method, and compares the performance of the second method to the performance of the first method.

\end{abstract}

\section{Introduction} 
\label{sec:Introduction}

Bentley introduced the \emph{k}-d tree as a binary search tree that stores \emph{k}-dimensional tuples \cite{Bentley}.  Similar to a binary search tree, a \emph{k}-d tree partitions a set of tuples at each recursive level of the tree.  Unlike a binary search tree that uses only one key for all levels of the tree, a \emph{k}-d tree uses $k$ keys and cycles through the keys for successive levels of the tree.  For example, to build a \emph{k}-d tree from three-dimensional tuples that represent $\left(x,y,z\right)$ coordinates, the keys would be cycled as $x,y,z,x,y,z...$  for successive levels of the \emph{k}-d tree. A more elaborate scheme for cycling the keys chooses the coordinate that has the widest dispersion or largest variance to be the key for a particular level of recursion \cite{Friedman}.

Bentley proposed that the $x$-, $y$-, and $z$-coordinates not be used as keys independently of one another, but instead that $x$, $y$, and $z$ form the most significant portions of the respective super keys $x$:$y$:$z$, $y$:$z$:$x$, and $z$:$x$:$y$ that represent cyclic permutations of $x$, $y$, and $z$.  The symbols for these super keys use a colon to designate the concatenation of the individual $x$, $y$ and $z$ values.  For example, the symbol $z$:$x$:$y$ represents a super key wherein $z$ is the most significant portion of the super key, $x$ is the middle portion of the super key, and $y$ is the least significant portion of the super key.

Willard proposed a dynamic \emph{k}-d tree composed as a set of static \emph{k}-d* trees, where the asterisk (*) designates that only the leaf nodes of the tree store \emph{k}-dimensional tuples \cite{Willard}.  Bentley and Saxe proposed set of static trees whose sizes are increasing integer powers of two \cite{Bentley79} \cite{Bentley80}, although not for a dynamic \emph{k}-d tree but instead for a dynamic Lipton-Tarjan data structure \cite{Lipton}. (Composing a dynamic tree as a set static trees whose sizes are increasing powers of two is now known as the \emph{logarithmic} method.) Overmars and van Leeuwen proposed an efficient deletion algorithm for the logarithmic method \cite{Overmars2}. Thereafter, they proposed rebuilding subtrees within a single dynamic \emph{k}-d* tree \cite{Overmars}.

All of these proposals are theoretical; none report implementations. Procopiuc et al. reported an implementation of a dynamic \emph{k}-d* tree built via the logarithmic method, wherein the leaf nodes are blocks of multiple \emph{k}-dimensional tuples \cite{Procopiuc}. Their implementation does not include the efficient deletion algorithm proposed by Overmars and van Leeuwen \cite{Overmars2}. The present article reports an implementation of a dynamic \emph{k}-d tree built via the logarithmic method, wherein \emph{k}-dimensional tuples are stored not only in leaf nodes of the static \emph{k}-d trees, but also in the internal nodes of those trees.

\section{Methods} 
\label{sec:Methods}

\begin{figure}[h]
\begin{mdframed}[backgroundcolor=white!255,linecolor=black!0]
\centering
\centerline{\includegraphics*[trim = {0.96in, 1.11in, 0.96in, 1.35in}, clip, width=\columnwidth]{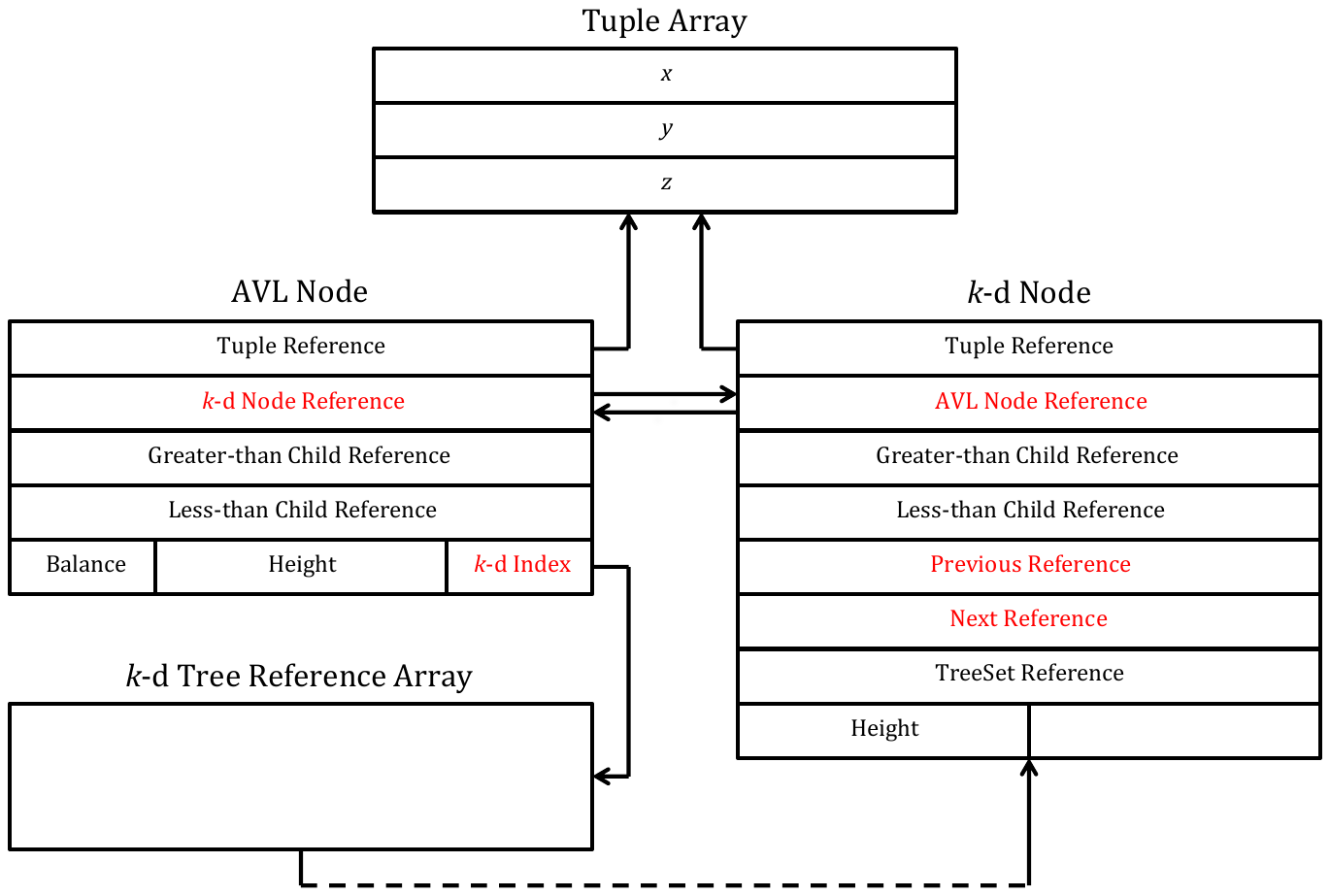}}
\caption{A logarithmic \emph{k}-d tree requires an AVL tree and a set of static \emph{k}-d trees.}
\label{fig:logarithmic_tree}
\end{mdframed}
\end{figure}

Figure \ref{fig:logarithmic_tree} depicts the tree-node structure required by a dynamic \emph{k}-d tree built via the logarithmic method, i.e., a logarithmic \emph{k}-d tree (in fact, a key-to-multiple-values map).

In Figure \ref{fig:logarithmic_tree}, each full-width field contains 64 integer bits. For example, in an AVL Node, the Tuple Reference field contains 64 bits, the Height field contains 32 bits, and the Balance and \emph{k}-d Index fields each contain 16 bits.

In Figure \ref{fig:logarithmic_tree}, the arrows represent references. An AVL Node references its associated \emph{k}-d Node via its \emph{k}-d Node Reference field, and a \emph{k}-d Node references its associated AVL Node via its AVL Node Reference field. The associated AVL Node and \emph{k}-d Node both reference the same Tuple Array via their Tuple Reference fields. 

As indicated by Figure \ref{fig:logarithmic_tree}, a logarithmic \emph{k}-d tree requires both an AVL tree \cite{Adelson} and a set of static \emph{k}-d trees. The AVL tree functions as a dictionary to facilitate deletion \cite{Overmars2}. In the absence of the AVL-tree dictionary, deletion of a tuple would require searching the first static tree, and then the second static tree, etc., until a \emph{k}-d node that references the tuple were found. However, the presence of the AVL-tree dictionary enables an initial search of that dictionary to find an AVL node that references the tuple. Then that AVL node's \emph{k}-d Index field is used to index the \emph{k}-d Tree Reference Array to obtain a reference to the root of the static \emph{k}-d tree that contains the tuple. Only that static tree is searched for the \emph{k}-d node that references the tuple. (The dashed arrow in Figure \ref{fig:logarithmic_tree} specifies that the \emph{k}-d Tree Reference Array references the \emph{k}-d Node indirectly via a search that begins at the root of the \emph{k}-d tree referenced by that array.)

As stated above, the logarithmic \emph{k}-d tree is in fact a key-to-multiple values map. This map requires the AVL-tree dictionary. In the absence of the AVL-tree dictionary, it would be possible to insert a key-value pair (using a tuple as the key) into one static tree, and then to use the same tuple as the key to insert the same key-value pair, or a different key-value pair, into another static tree. The AVL-tree dictionary prevents this kind of redundant insertion, because an initial search of the AVL tree attempts to find an AVL node that references the tuple. If an AVL node is found, its \emph{k}-d Node Reference field is used to obtain the associated \emph{k}-d node that also references the tuple, and then that \emph{k}-d node's TreeSet Reference field is used to obtain the \lstinline{TreeSet} into which to insert the value. But if an AVL node is not found, then an AVL node is created, and the logarithmic \emph{k}-d tree insertion algorithm described below is invoked.

\subsection{Insertion Algorithm for a Logarithmic \emph{k}-d Tree}
\label{sec:Insertion}

A logarithmic \emph{k}-d tree comprises a set of static \emph{k}-d trees $\mathrm{T}_i$ whose sizes $\left | \mathrm{T}_i \right |$ are constrained as follows unless the tree is empty, in which case $\left | \mathrm{T}_i \right | = 0$.
\begin{equation}
2^{i-2} < \left | \mathrm{T}_i \right | \leq 2^i
\label{eq:size}
\end{equation}
The non-inclusive lower limit $2^{i-2}$ is imposed by the \emph{k}-d tree efficient deletion algorithm, as will be explained later \cite{Overmars2}.

$2^{i-2}$ is undefined for $i < 2$ in an integer space. For this reason, $\mathrm{T}_0$ and $\mathrm{T}_1$ are special cases. Non-empty $\mathrm{T}_0$ contains 1 tuple. Non-empty $\mathrm{T}_1$ contains 1 or 2 tuples.

Insertion of a key-value pair (in fact, a super-key-value pair) into a logarithmic \emph{k}-d tree begins with a search for the super key, i.e., the tuple, in the AVL tree. If an AVL node is found that references the tuple, that AVL node's \emph{k}-d Node Reference field is used to obtain a \emph{k}-d node whose TreeSet Reference field is used to obtain a TreeSet into which to insert the value. If the tuple is not found, a new AVL node is inserted into the AVL tree, and a new static \emph{k}-d tree is built to contain a new \emph{k}-d node.

Creation of the new static \emph{k}-d tree begins by iterating over the set of static \emph{k}-d trees, beginning with $\mathrm{T}_0$, to find the first empty tree $\mathrm{T_i}$. As that scan proceeds, the sum $s$ of static tree sizes,
\begin{equation}
s = \sum_{j=0}^{i-1}\left | \mathrm{T}_j \right |
\label{eq:sum}
\end{equation}
is accumulated for trees $\mathrm{T}_0 \cdots \mathrm{T}_{i-1}$ \cite{Overmars2}.  The \emph{k}-d nodes are harvested from trees $\mathrm{T}_0 \cdots \mathrm{T}_{i-1}$ for re-use in building the new static \emph{k}-d tree. Harvesting is facilitated by each node's Previous Reference and Next Reference fields (depicted in Figure \ref{fig:logarithmic_tree}) that maintain a doubly linked list of the nodes that belong to a particular static tree. The linked lists from trees $\mathrm{T}_0 \cdots \mathrm{T}_{i-1}$ are catenated to form a list all their nodes, and then a new \emph{k}-d node that references the inserted super-key-value pair is appended to that list. The list is provided to an $O \left [ n \log \left ( n \right ) \right ]$ tree-building algorithm that builds the new static \emph{k}-d tree \cite{Brown2015}, and that destroys trees $\mathrm{T}_0 \cdots \mathrm{T}_{i-1}$ by re-assigning each \emph{k}-d node's Greater-than Child Reference and Less-than Child Reference fields (depicted in Figure \ref{fig:logarithmic_tree}). If the sum of static tree sizes $s \le 2^{i-1}$, the new static \emph{k}-d tree is $\mathrm{T}_{i-1}$. But if $s > 2^{i-1}$, the new static \emph{k}-d tree is $\mathrm{T}_i$. Assignment of the new static \emph{k}-d tree to either $\mathrm{T}_{i-1}$ or $\mathrm{T}_i$ is accomplished by iterating over the new static tree's doubly linked list of \emph{k}-d nodes, obtaining an AVL node using each \emph{k}-d node's AVL Node Reference field, and setting that AVL node's \emph{k}-d Index field to either $i {-} 1$ or $i$ respectively \cite{Overmars2}.

\subsection{Efficient Deletion Algorithm for a Logarithmic \emph{k}-d Tree}
\label{sec:Deletion}

Deletion of a super-key-value pair from a logarithmic \emph{k}-d tree begins with a search for the super key, i.e., the tuple, in the AVL tree. If an AVL node is found that references the tuple, that AVL node's \emph{k}-d Node Reference field is used to obtain a \emph{k}-d node whose TreeSet Reference field is used to obtain a TreeSet from which to erase the value. If erasure produces an empty TreeSet, the AVL node is deleted from the AVL tree, and the \emph{k}-d node is deleted from the static \emph{k}-d tree $\mathrm{T}_i$ specified by the AVL node's \emph{k}-d Index field. Deletion is accomplished by the deletion algorithm for a single, dynamic \emph{k}-d tree \cite{Brown2026}. Rebalancing that tree is optional, as will be discussed later.

After deletion of the \emph{k}-d node from static \emph{k}-d tree $\mathrm{T}_i$, it may be necessary to rebuild $\mathrm{T}_i$, $\mathrm{T}_{i-1}$, or $\mathrm{T}_{i-2}$, depending on the result of the following sequence of tests.

1. If $\left | \mathrm{T}_i \right | > 2^{i-2}$, rebuilding is unnecessary because $\mathrm{T}_i$ satisfies the constriants of Equation \ref{eq:size}. But if $\left | \mathrm{T}_i \right | \ngtr 2^{i-2}$, rebuilding may be necessary, and $\left | \mathrm{T}_i \right | = 2^{i-2}$ because $\mathrm{T}_i$ satisfied the inequality $\left | \mathrm{T}_i \right | > 2^{i-2}$ prior to deletion of the \emph{k}-d node.

2. If $\left | \mathrm{T}_i \right | = 2^{i-2}$ and $\left | \mathrm{T}_{i-1} \right | \neq 0$, then the \emph{k}-d nodes from $\mathrm{T}_{i-1}$ and $\mathrm{T}_i$ are harvested and re-used to build a new static \emph{k}-d tree via the $O \left [ n \log \left ( n \right ) \right ]$ tree-building algorithm, which destroys trees $\mathrm{T}_{i-1}$ and $\mathrm{T}_i$ while building the new static tree. Prior to destruction of these trees, $\left | \mathrm{T}_{i-1} \right |$ is tested to determine whether to assign the new static tree to $\mathrm{T}_{i-1}$ or to $\mathrm{T}_i$. If $\left | \mathrm{T}_{i-1} \right | \leq 2^{i-2}$, the following inequality applies
\begin{equation}
\left | \mathrm{T}_{i-1} \right | + \left | \mathrm{T}_i \right | \leq 2*2^{i-2} \ .
\label{eq:T1less2}
\end{equation}
This inequality is equivalent to
\begin{equation}
\left | \mathrm{T}_{i-1} \right | + \left | \mathrm{T}_i \right | \leq 2^{i-1} 
\label{eq:T1less1}
\end{equation}
so the new static tree is assigned to $\mathrm{T}_{i-1}$, consistent with Equation \ref{eq:size} applied to $\mathrm{T}_{i-1}$. But if $\left | \mathrm{T}_{i-1} \right | > 2^{i-2}$, the following inequality applies
\begin{equation}
\left | \mathrm{T}_{i-1} \right | + \left | \mathrm{T}_i \right | > 2^{i-1}
\label{eq:T1greater}
\end{equation}
so the new static tree is assigned to $\mathrm{T}_i$, consistent with Equation \ref{eq:size} applied to $\mathrm{T}_i$. Assignment of the new static \emph{k}-d tree to either $\mathrm{T}_{i-1}$ or $\mathrm{T}_i$ is accomplished by iterating over the new static tree's doubly linked list of \emph{k}-d nodes, obtaining an AVL node using each \emph{k}-d node's AVL Node Reference field, and setting that AVL node's \emph{k}-d Index field to either $i {-} 1$ or $i$ respectively \cite{Overmars2}.

3. If $\left | \mathrm{T}_i \right | = 2^{i-2}$ and $\left | \mathrm{T}_{i-1} \right | = 0$ and $\left | \mathrm{T}_{i-2} \right | \neq 0$, the following inequality applies
\begin{equation}
\left | \mathrm{T}_{i-2} \right | + \left | \mathrm{T}_i \right | \leq 2*2^{i-2} \ .
\label{eq:T2less2}
\end{equation}
This inequality is equivalent to
\begin{equation}
\left | \mathrm{T}_{i-2} \right | + \left | \mathrm{T}_i \right | \leq 2^{i-1} \ .
\label{eq:T2less1}
\end{equation}
In this case, the \emph{k}-d nodes from $\mathrm{T}_{i-2}$ and $\mathrm{T}_i$ are harvested and re-used to build a new static \emph{k}-d tree via the $O \left [ n \log \left ( n \right ) \right ]$ tree-building algorithm, which destroys trees $\mathrm{T}_{i-2}$ and $\mathrm{T}_i$ during the process of building the new static tree. The new static tree is assigned to $\mathrm{T}_{i-1}$, consistent with Equation \ref{eq:size} applied to $\mathrm{T}_{i-1}$. This assignment is accomplished by iterating over the new static tree's doubly linked list of \emph{k}-d nodes, obtaining an AVL node using each \emph{k}-d node's AVL Node Reference field, and setting that AVL node's \emph{k}-d Index field to $i {-} 1$.

4. If $\left | \mathrm{T}_i \right | = 2^{i-2}$ and $\left | \mathrm{T}_{i-1} \right | = 0$ and $\left | \mathrm{T}_{i-2} \right | = 0$, then $\mathrm{T}_{i-2}$ must be swapped with $\mathrm{T}_i$ because $\left | \mathrm{T}_i \right | $ satisfies the Equation \ref{eq:size} applied to $\mathrm{T}_{i-2}$. It is not necessary to build a new static \emph{k}-tree. The swap is accomplished by swapping the references at indices $i$ and $i {-} 2$ in the \emph{k}-d Tree Reference Array depicted in Figure \ref{fig:logarithmic_tree}. After this swap, the \emph{k}-d Tree Reference array elements at indices $i$ and $i {-} 2$ reference respectively $\mathrm{T}_{i-2}$ (for which $\left | \mathrm{T}_{i-2} \right | = 0$) and $\mathrm{T}_i$ (for which $\left | \mathrm{T}_i \right | = 2^{i-2}$). The final step of the swap is accomplished by iterating over $\mathrm{T}_i$'s doubly linked list of \emph{k}-d nodes, obtaining an AVL node using each \emph{k}-d node's AVL Node Reference field, and setting that AVL node's \emph{k}-d Index field to $i {-} 2$.

\subsection{Rebalancing After Deletion from a Logarithmic \emph{k}-d Tree}
\label{sec:Rebalancing}

It is possible to rebalance a single, static \emph{k}-d tree after deletion of a \emph{k}-node using a previously described rebalancing algorithm \cite{Brown2026}. However, this rebalancing is optional because the single, static \emph{k}-d tree was built initially as a balanced tree via a static \emph{k}-d tree-building algorithm \cite{Brown2015}, and because deletion can only decrease, not increase, the depth of that tree.

Moreover, rebalancing is not required for cases 2 and 3 described in Section \ref{sec:Deletion} because these cases rebuild via the static \emph{k}-d tree-building algorithm that builds a balanced tree. On the other hand, rebalancing might improve performance for cases 1 and 4 that do not utilize the static \emph{k}-d tree-building algorithm.

In the benchmarks reported below, rebalancing is either not performed after deletion, or performed after deletion for cases 1 through 4, or performed after deletion for only cases 1 and 4, in order to assess the benefit of optional rebalancing.

\subsection{Improved Insertion Algorithm for a Logarithmic \emph{k}-d Tree}
\label{sec:Improved_Insertion}

The efficient deletion algorithm for a logarithmic \emph{k}-d tree produces non-full trees within the set of static \emph{k}-d trees. However, the insertion algorithm for a logarithmic \emph{k}-d tree cannot insert into these non-full trees because it scans only for an empty tree. Hence, for a dynamic environment wherein deletion and insertion are interleaved, an improved insertion algorithm that inserts into non-full trees is necessary.

Similar to the insertion algorithm described in Section \ref{sec:Insertion}, the improved insertion algorithm begins with a search for the super key, i.e., the tuple, in the AVL tree. If an AVL node is found that references the tuple, that AVL node's \emph{k}-d Node Reference field is used to obtain a \emph{k}-d node whose TreeSet Reference field is used to obtain a TreeSet into which to insert the value associated with the super key. 

If the tuple is not found, a new AVL node is inserted into the AVL tree. Then the improved insertion algorithm begins with $\mathrm{T}_0$ but does not scan for the first empty tree whose size $\left | \mathrm{T}_i \right | = 0$. Instead, it scans for the first non-full tree $\mathrm{T}_i$ whose size $\left | \mathrm{T}_i \right |$ is constrained as follows.
\begin{equation}
2^{i-2} < \left | \mathrm{T}_i \right | < 2^i 
\label{eq:modiffied_size}
\end{equation}
If a non-full tree is found, the improved insertion algorithm inserts a new \emph{k}-d node into that tree via the insertion algorithm for a single, dynamic \emph{k}-d tree, and rebalances the tree \cite{Brown2026}. On the other hand, if a non-full tree is not found, the improved insertion algorithm reverts to the \emph{unimproved} insertion algorithm described in Section \ref{sec:Insertion} that scans for the first empty tree.

\subsection{Benchmark Methodology}

To assess the performance of the dynamic \emph{k}-d tree, benchmarks were executed on a Hewlett-Packard Pro Mini 400 G9 with 2x32GB DDR5-4800 RAM and a 14th-generation Intel Raptor Lake CPU (i7 14700T with 8 performance cores, 5.2GHz performance core maximum frequency, 78.6GB/s maximum memory bandwidth, 80KB per-core L1 and 2MB per-core L2 caches, and a 33MB L3 cache shared by all cores).

A \emph{k}-d tree benchmark measured algorithm performance for trees that contained $n$ nodes, for values of $n$ in the range $ \left[ 1,003,201; 4,523,071 \right] $ that map to equally spaced values of $n \log_2 \left( n \right)$ in the range $\left[ 20,000,000; 100,000,000 \right]$. Each node of the tree stored a $k$-dimensional tuple of 64-bit integers. The integers were equally spaced across the maximum 64-bit integer range $r$, so the spacing was $r/n$. The integers were randomly shuffled via the Fisher-Yates algorithm \cite{Fischer} using the \lstinline{java.util.Random} random-number generator, and copied to the first of the $k$ dimensions, then randomly shuffled again and copied to the second of the $k$ dimensions, et cetera. So that all benchmarks randomly shuffled the integers in an identical sequence, each benchmark initialized the random-number generator to the same seed via the \lstinline{rand.setSeed} method.

The \emph{k}-d tree benchmark was implemented in Java, compiled via javac 21.0.11, and executed under Ubuntu 24.04.1 LTS via a single thread mapped to a single performance core via the Ubuntu \lstinline{taskset} command.

Each benchmark measured execution times, via the \lstinline{System.currentTimeMillis} method, for insertion of all tuples into the tree via the unimproved insertion algorithm (Section \ref{sec:Insertion}) and for deletion of all tuples from the tree without rebalancing (Section \ref{sec:Deletion}). To provide a comparison, the execution time was measured for insertion and deletion via the dynamic \emph{k}-d tree algorithms \cite{Brown2026}. Each benchmark was repeated 100 times and the mean values and standard deviations of the execution times were calculated. The standard deviations were less than 10\% of the mean values.

\subsection{Benchmark Results}

\begin{figure}[h]
\centering
\centerline{\includegraphics*[trim = {1.00in, 3.47in, 1.33In, 1.52In}, clip, width=\columnwidth]{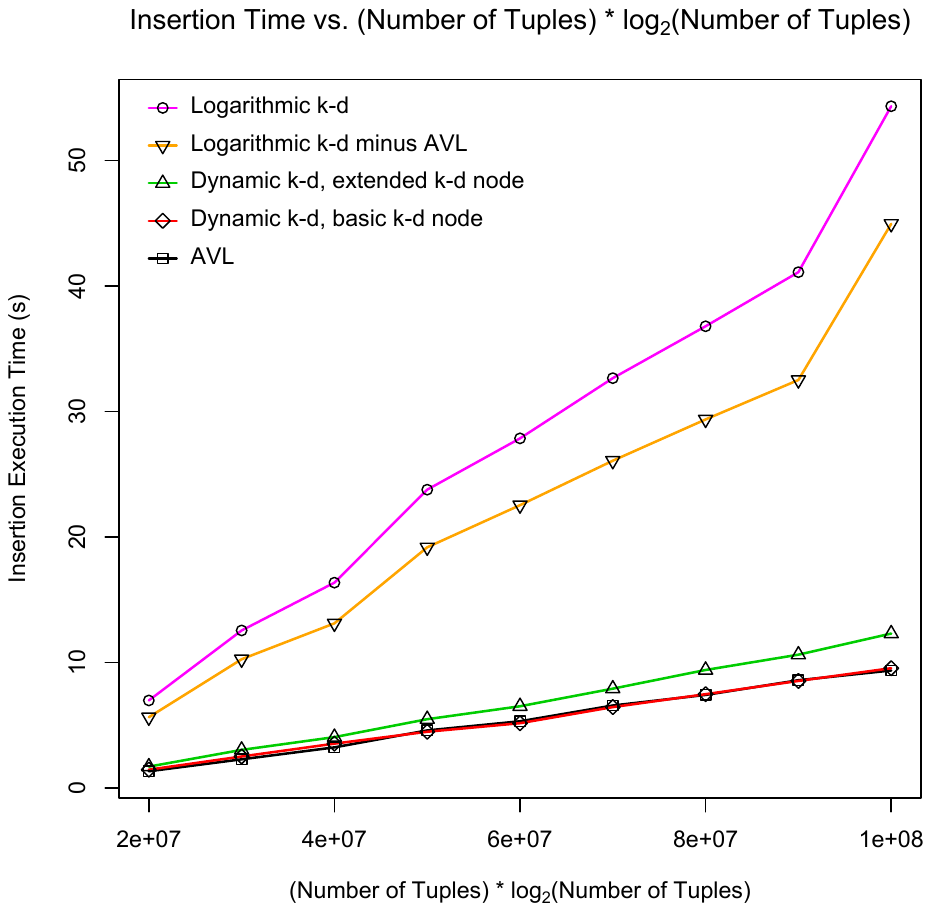}}
\caption{Insertion times for logarithmic and dynamic \emph{k}-d trees and an AVL tree}
\label{fig:insertion}
\end{figure}

Figure \ref{fig:insertion} plots the amortized execution times (in seconds) for single-threaded insertion into a logarithmic \emph{k}-d tree, an AVL tree, and two dynamic \emph{k}-d trees. All execution times are plotted versus $n \log_2 \left( n \right)$ where $n$ represents the number of tuples in the tree. One dynamic \emph{k}-d tree comprises \emph{basic} \emph{k}-d nodes whose fields are labeled in black in Figure \ref{fig:logarithmic_tree}. The other dynamic \emph{k}-d tree comprises \emph{extended} \emph{k}-d nodes that are required by the logarithmic \emph{k}-d tree, and whose fields are labeled in black and red in Figure \ref{fig:logarithmic_tree}.

Figure \ref{fig:insertion} reveals that the insertion time for the dynamic \emph{k}-d tree that comprises extended \emph{k}-d nodes exceeds the insertion time for the dynamic \emph{k}-d tree that comprises basic \emph{k}-d nodes. This difference in insertion times may be explained by the fact that an extended \emph{k}-d node is larger than a basic \emph{k}-d node and hence requires more memory operations to read or write.

Figure \ref{fig:insertion} also plots the insertion time for the logarithmic \emph{k}-d tree, as well as the insertion time for the logarithmic \emph{k}-d tree minus the insertion time for the AVL tree. Subtracting the AVL tree insertion time from the logarithmic \emph{k}-d tree insertion time reveals that the increased insertion time for the logarithmic \emph{k}-d tree relative to the insertion time for either dynamic \emph{k}-d tree cannot be fully explained by AVL-tree operations associated with insertion into the logarithmic \emph{k}-d tree. 

\begin{figure}[h]
\centering
\centerline{\includegraphics*[trim = {1.00in, 3.47in, 1.33In, 1.52In}, clip, width=\columnwidth]{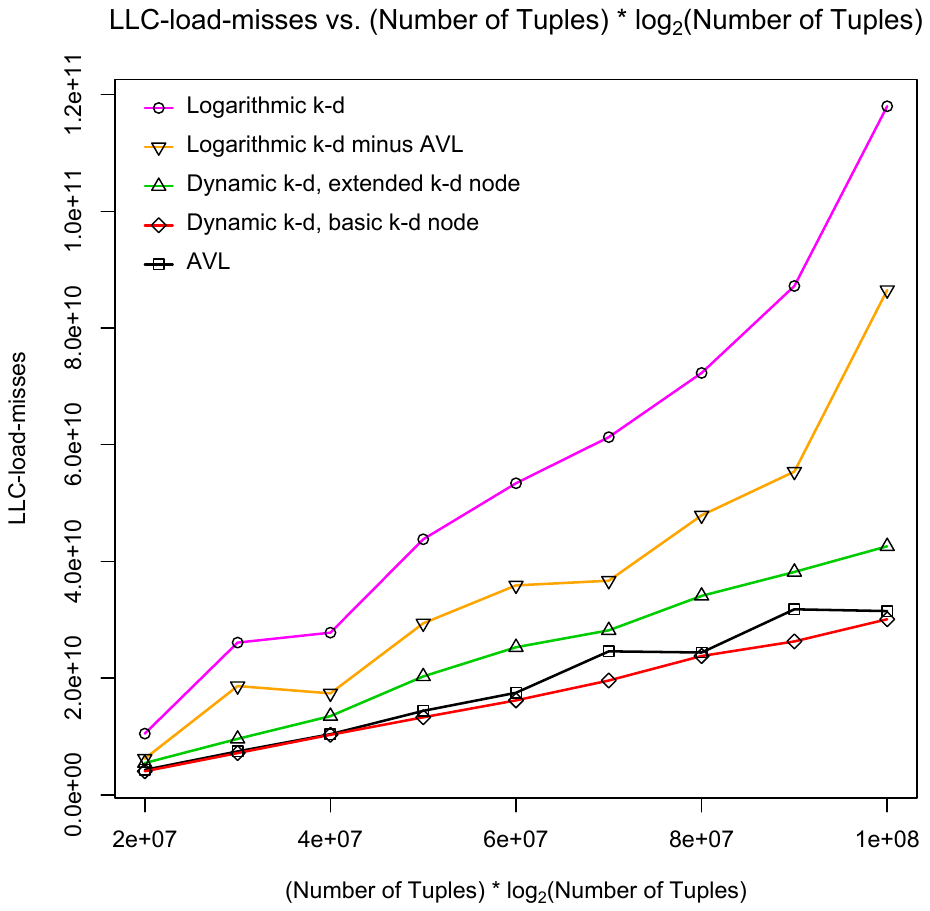}}
\caption{LLC-load-misses for insertion into and deletion from various trees}
\label{fig:cache}
\end{figure}

An additional explanation recognizes that insertion requires building a static \emph{k}-d tree and then iterating over that tree's doubly linked list of \emph{k}-d nodes to obtain an AVL node using each \emph{k}-d node's AVL Node Reference field and to assign a value to that AVL node's \emph{k}-d Index field. This iteration causes cache-line misses, as measured via the Ubuntu \lstinline{perf stat -e LLC-load-misses} command. Figure \ref{fig:cache} reveals that the combined LLC-load-misses for insertion into and deletion from the logarithmic \emph{k}-d tree minus the combined LLC-load-misses for the AVL tree exceeds the combined LLC-load-misses for either dynamic \emph{k}-d tree. Hence, cache utilization is less efficient for the logarithmic \emph{k}-d tree than for either dynamic \emph{k}-d tree.

 \newpage
 
\begin{figure}[h]
\centering
\centerline{\includegraphics*[trim = {1.00in, 3.47in, 1.33In, 1.52In}, clip, width=\columnwidth]{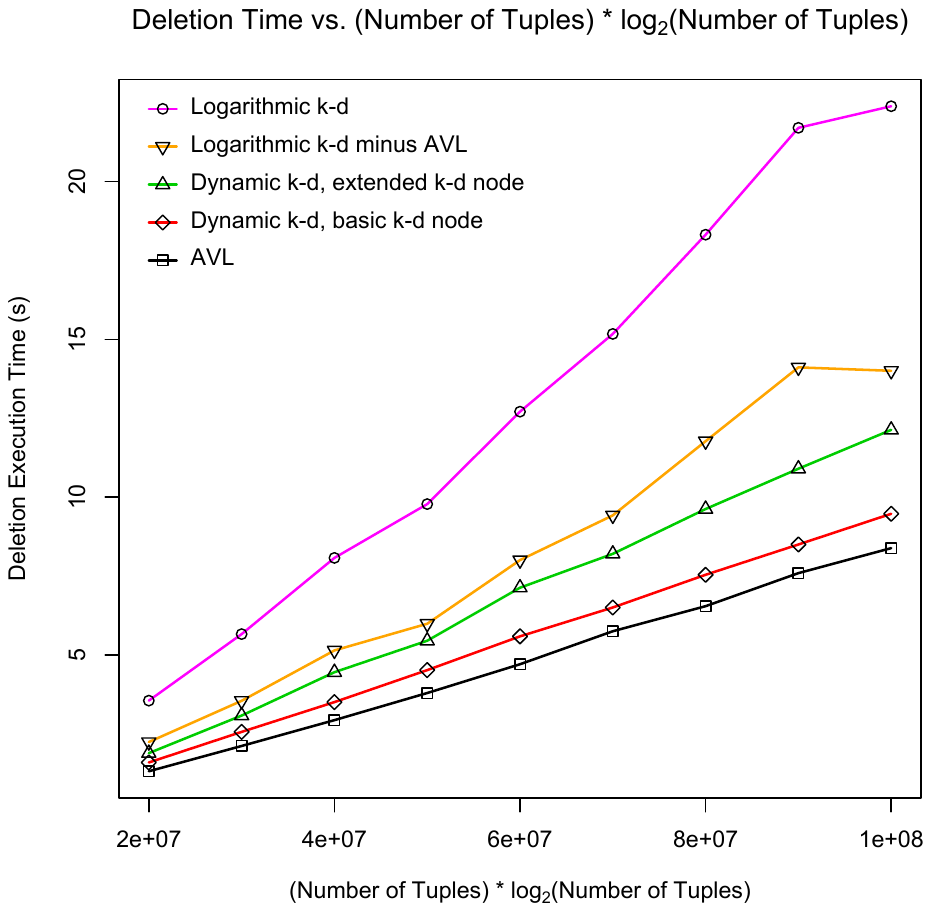}}
\caption{Deletion times for logarithmic and dynamic \emph{k}-d trees and an AVL tree}
\label{fig:deletion}
\end{figure}

Figure \ref{fig:deletion} plots amortized execution times (in seconds) for single-threaded deletion from a logarithmic \emph{k}-d tree, an AVL tree, and two dynamic \emph{k}-d trees. All execution times are plotted versus $n \log_2 \left( n \right)$ where $n$ represents the number of tuples in the tree. One dynamic \emph{k}-d tree comprises basic \emph{k}-d nodes whose fields are labeled in black in Figure \ref{fig:logarithmic_tree}. The other dynamic \emph{k}-d tree comprises extended \emph{k}-d nodes that are required by the logarithmic \emph{k}-d tree, and whose fields are labeled in black and red in Figure \ref{fig:logarithmic_tree}.

Figure \ref{fig:deletion} reveals that the deletion time for the dynamic \emph{k}-d tree that comprises extended \emph{k}-d nodes exceeds the deletion time for the dynamic \emph{k}-d tree that comprises basic \emph{k}-d nodes. This difference in deletion times may be explained by the fact that an extended \emph{k}-d node is larger than a basic \emph{k}-d node and hence requires more memory operations to read or write.

Figure \ref{fig:deletion} also plots the deletion time for the logarithmic \emph{k}-d tree, as well as the deletion time for the logarithmic \emph{k}-d tree minus the deletion time for the AVL tree. Subtracting the AVL tree deletion time from the logarithmic \emph{k}-d tree deletion time reveals that the increased deletion time for the logarithmic \emph{k}-d tree relative to the deletion time for either dynamic \emph{k}-d tree cannot be fully explained by AVL-tree operations associated with deletion from the logarithmic \emph{k}-d tree. An additional explanation has been suggested by the above discussion of Figure \ref{fig:cache}.

Figures \ref{fig:insertion} and \ref{fig:deletion} report execution times that were obtained first by inserting all of the super-key-value pairs into logarithmic and dynamic \emph{k}-d trees, and then by deleting all of the super-key-value pairs from those trees. This approach does not mimic a dynamic environment wherein deletion and insertion are interleaved. To mimic a dynamic environment, a second series of benchmarks was performed.

For this second series, the super-key-value pairs were subdivided into eight segments. First, all eight segments were inserted into logarithmic and dynamic \emph{k}-d trees. Then the pairs within the first segment were randomly shuffled and deleted from the trees. Then these pairs were randomly shuffled and re-inserted into the trees. Then the second through eighth segments were processed sequentially in the same manner.

\begin{figure}[h]
\centering
\centerline{\includegraphics*[trim = {1.00in, 3.47in, 1.33In, 1.52In}, clip, width=\columnwidth]{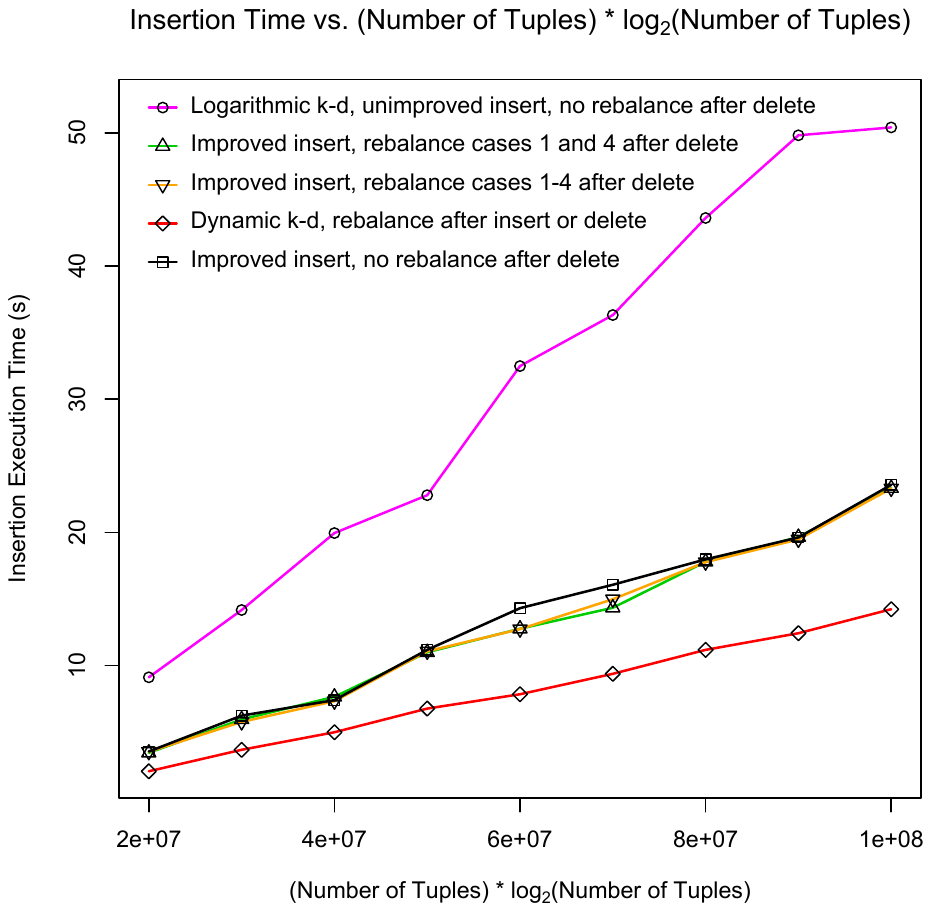}}
\caption{Segmented insertion times for logarithmic and dynamic \emph{k}-d trees}
\label{fig:segmented_insert}
\end{figure}

Figure \ref{fig:segmented_insert} plots the amortized execution times (in seconds) for single-threaded, segmented insertion (preceded by segmented deletion) into a logarithmic \emph{k}-d tree and into a dynamic \emph{k}-d tree that comprises extended \emph{k}-d nodes.

In Figure \ref{fig:segmented_insert}, the magenta curve plots the execution time for unimproved insertion of segments into a logarithmic \emph{k}-d tree (Section \ref{sec:Insertion}) that were previously deleted without rebalancing (Section \ref{sec:Deletion}). The black, orange, and green curves plot the execution time for improved insertion of segments into a logarithmic \emph{k}-d tree (Section \ref{sec:Improved_Insertion}) that were previously deleted respectively without rebalancing, with rebalancing for cases 1 through 4, and with rebalancing for only cases 1 and 4 (Section \ref{sec:Rebalancing}). The red curve plots the execution time for insertion of segments into a dynamic \emph{k}-d tree that were previously deleted, where both insertion and deletion were followed by rebalancing. These curves demonstrate that rebalancing associated with deletion does not significantly impact the insertion time. These curves also demonstrate that, although improved insertion does decrease the insertion time for a logarithmic \emph{k}-d tree, that decreased insertion time remains significantly longer than the insertion time for a dynamic \emph{k}-d tree.

\begin{figure}[h]
\centering
\centerline{\includegraphics*[trim = {1.00in, 3.47in, 1.33In, 1.52In}, clip, width=\columnwidth]{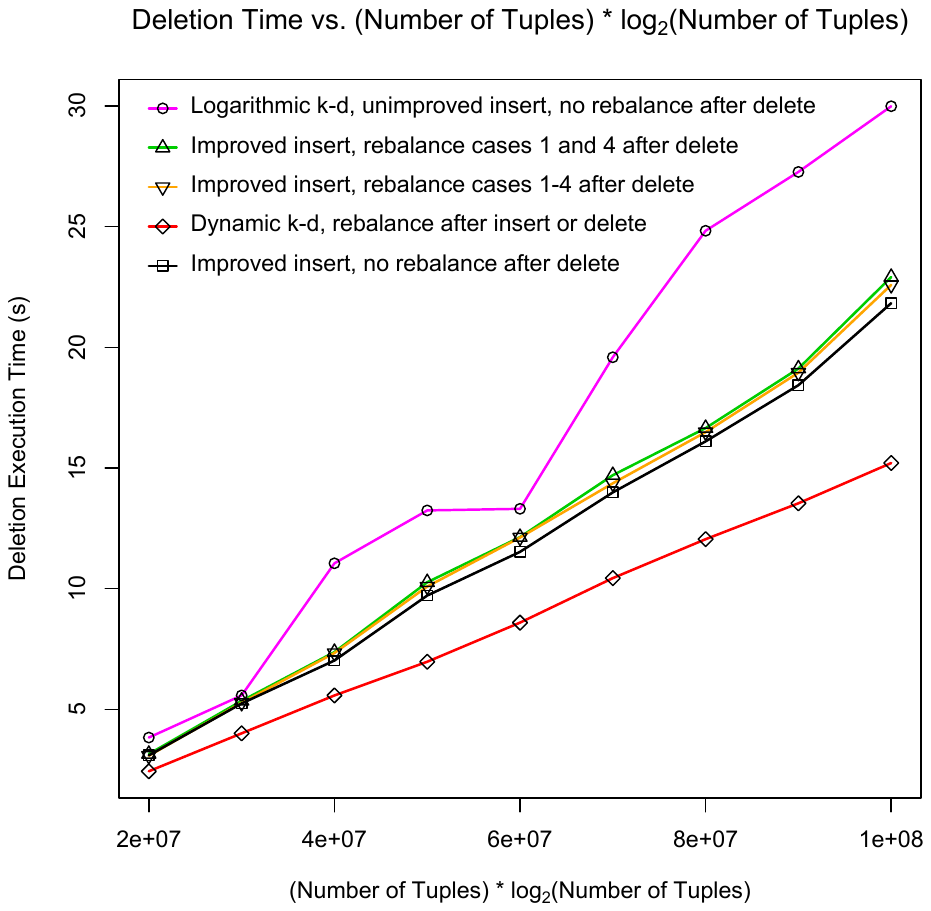}}
\caption{Segmented deletion times for logarithmic and dynamic \emph{k}-d trees}
\label{fig:segmented_delete}
\end{figure}

Figure \ref{fig:segmented_delete} plots the amortized execution times (in seconds) for single-threaded, segmented deletion (followed by segmented insertion) from a logarithmic \emph{k}-d tree and from a dynamic \emph{k}-d tree that comprises extended \emph{k}-d nodes.

In Figure \ref{fig:segmented_delete}, the magenta curve plots the execution time for deletion of segments from a logarithmic \emph{k}-d tree without rebalancing (Section \ref{sec:Deletion}) that were subsequently inserted via unimproved insertion (Section \ref{sec:Insertion}). The black, orange, and green curves plot the execution time for deletion of segments from a logarithmic \emph{k}-d tree respectively without rebalancing, with rebalancing for cases 1 through 4, and with rebalancing for only cases 1 and 4 (Section \ref{sec:Rebalancing}), where those segments were subsequently inserted via improved insertion (Section \ref{sec:Improved_Insertion}). The red curve plots the execution time for deletion of segments from a dynamic \emph{k}-d tree that were subsequently inserted, where both deletion and insertion were followed by rebalancing. These curves demonstrate that rebalancing associated with deletion does not significantly impact the deletion time. These curves also demonstrate that, although improved insertion does decrease the deletion time for a logarithmic \emph{k}-d tree, that decreased deletion time remains significantly longer than the deletion time for a dynamic \emph{k}-d tree.

\section{Conclusions}
\label{Conclusions}

The logarithmic \emph{k}-d tree requires a more elaborate data structure than the dynamic \emph{k}-d tree and it exhibits significantly lower performance than the dynamic \emph{k}-d tree. The lower performance is due to AVL-tree operations associated with insertion and deletion, and due to the requirement to iterate over a doubly linked list of \emph{k}-d nodes when building a static tree during insertion and deletion. Both the associated AVL-tree operations and the iteration cause cache LLC-load-misses that decrease performance.

Future research could explore whether a dynamic \emph{k}-d* tree outperforms a dynamic \emph{k}-d tree, where the asterisk (*) designates that only the tree's leaf nodes store super-key-value pairs. Deletion from a dynamic \emph{k}-d or \emph{k}-d* tree of height $h$ exhibits respectively $O \left [ 2^ { h \left ( 1-1/k \right ) } \right ]$ or $O \left ( h \right ) $ computational complexity \cite{Willard}.

\section*{Supplemental Materials}

Included with this manuscript are Java implementations of $O \left[ n \log \left( n \right) \right]$ and $O \left[ kn \log \left( n \right) \right]$ algorithms that insert into and delete from logarithmic, dynamic, and static \emph{k}-d tree-based key-to-multiple-values maps. These implementations are also available at

\smallskip
\href{https://github.com/RussellABrown/kd-tree}{https://github.com/RussellABrown/kd-tree}

\section*{Author Contact Information}

\href{https://www.linkedin.com/in/russellabrown/}{https://www.linkedin.com/in/russellabrown/}


\section*{References}

\small
\bibliographystyle{jcgt}
\bibliography{paper}

@ARTICLE{Bentley,
  author = {J.L. Bentley},
  title = {Multidimensional binary search trees used for associative searching},
  journal = {Communications of the ACM},
  volume = {18},
  issue = {9},
  pages = "509-517",
  year = 1975,
  url = {https://dl.acm.org/toc/cacm/1975/18/9},
  doi = {10.1145/361002.361007}
}

@ARTICLE{Friedman,
  author = {J.H. Friedman and J.L. Bentley and R.A. Finkel},
  title = {An algorithm for finding best matches in logarithmic expected time},
  journal = {ACM Transactions on Mathematical Software},
  volume = {3},
  issue = {3},
  pages = "209-226",
  year = 1977,
  url = {http://dl.acm.org/citation.cfm?id=355745},
  doi = {10.1145/355744.355745}
}

@ARTICLE{Bentley79,
  author = {J.L. Bentley},
  title = {Decomposable searching problems},
  journal = {Information Processing Letters},
  volume = {8},
  issue = {5},
  pages = "244-251",
  year = 1979,
  url = {https://www.sciencedirect.com/science/article/abs/pii/0020019079901170},
  doi = {10.1016/0020-0190(79)90117-0}
}

@ARTICLE{Bentley80,
  author = {J.L. Bentley and J.B. Saxe},
  title = {Decomposable searching problems {I}. {S}tatic-to-dynamic transformation},
  journal = {Journal of Algorithms},
  volume = {1},
  issue = {4},
  pages = "301-358",
  year = 1980,
  url = {https://www.sciencedirect.com/science/article/abs/pii/0196677480900152},
  doi = {10.1016/0196-6774(80)90015-2}
}

@ARTICLE{Lipton,
  author = {R.J. Lipton and R.E. Tarjan},
  title = {Applications of a planar separator theorum},
  journal = {SIAM Journal on Computing},
  volume = {9},
  issue = {3},
  pages = "615-627",
  year = 1980,
  url = {https://epubs.siam.org/doi/10.1137/0209046},
  doi = {10.1137/0209046}
}

@ARTICLE{Overmars2,
  author = {M.H. Overmars and J. {van Leeuwen}},
  title = {Two general methods for dynamizing decomposable searching problems},
  journal = {Computing},
  volume = {26},
  pages = "155-166",
  year = 1981,
  url = {https://link.springer.com/article/10.1007/BF02241781},
  doi = {10.1007/BF02241781}
}

@ARTICLE{Overmars,
  author = {M.H. Overmars and J. {van Leeuwen}},
  title = {Dynamic multi-dimensional data structures based on quad- and k-d trees},
  journal = {Acta Informatica},
  volume = {17},
  pages = "267-285",
  year = 1982,
  url = {https://dl.acm.org/doi/abs/10.1007/BF00264354},
  doi = {10.1007/BF00264354}
}

@INCOLLECTION{Procopiuc,
   author = {O. Procopiuc and P.K. Agarwal and L. Arge and J.S. Vittner},
   title = {Bkd-tree: {A} dynamic scalable kd-tree},
   booktitle = {Lecture Notes in Computer Science (LNCS), Advances in Spatial and Temporal Databases}, 
   editor = {T. Hadzilacos and Y. Manolopoulos and J. Roddick and Y. Theodoridis},
   volume = {2750},
   pages = "46-65", 
   year = 2003, 
   publisher = {Springer-Verlag},
   address = {Berlin},
   url = {https://doi.org/10.1007/978-3-540-45072-6\_4},
  doi = {10.1007/978-3-540-45072-6\_4}
}

@ARTICLE{Adelson,
   AUTHOR  = {G.M. Adelson-Velskii and E.M. Landis},
   TITLE   = {An algorithm for the organization of information},  
   YEAR    = {1962},
   JOURNAL = {Soviet Mathematics Doklady},
   VOLUME  = {3},
   PAGES   = {1259-1263},
   URL = {https://zhjwpku.com/assets/pdf/AED2-10-avl-paper.pdf}
}

@ARTICLE{Brown2015,
  author = {R.A. Brown},
  title = {Building a balanced \emph{k}-d tree in \emph{{O}}(\emph{kn} log \emph{n}) time},
  journal = {Journal of Computer and Graphics Techniques (JCGT)},
  volume = {7},
  pages = "50-68",
  year = 2015,
  url = {http://jcgt.org/published/004/01/03/}
}

@ARTICLE{Brown2026,
  author = {R.A. Brown},
  title = {A dynamic, self-balancing \emph{k}-d tree},
  journal = {arXiv:2509.08148 [cs.DS]},
  pages = "1-19",
  year = 2026,
  url = {https://arxiv.org/abs/2509.08148},
  doi = {10.48550/arXiv.2509.08148}
}

@INCOLLECTION{Willard,
   AUTHOR  = {D.E. Willard},
   TITLE   = {Balanced forests of k-d* trees as a dynamic data structure},  
   YEAR    = {1978},
   BOOKTITLE = {Aiken Computation Lab TR-23-78},
   PAGES   = {1-29},
   PUBLISHER = {Harvard University, Cambridge, MA},
   URL = {https://apps.dtic.mil/sti/pdfs/ADA110403.pdf}
}

@misc{Fischer,
    author= {Wikipedia},
    year = {2026},
    title = {Fischer-{Y}ates shuffle},
    note  = {\url{https://en.wikipedia.org/wiki/Fisher\%E2\%80\%93Yates_shuffle}, 
             Last accessed on 2026-07-21},
}

\end{document}